\documentclass[amssymb]{revtex4}
\usepackage{amssymb}
\usepackage{graphicx}
\usepackage{color}
\usepackage{textcomp}
\usepackage{ulem}

\begin{document}
\title{Dynamically generated magnetic moment in the Wigner-function formalism}
\author{Shijun Mao$^{1,2}$ and Dirk H.\ Rischke$^{2,3}$}
\affiliation{${}^1$School of Science, Xi'an Jiaotong University, Xi'an, Shaanxi 710049, China\\
${}^2$Institute for Theoretical Physics, Goethe University, Max-von-Laue-Str.\ 1, D-60438 Frankfurt am Main, Germany\\
${}^3$Interdisciplinary Center for Theoretical Study and Department of Modern Physics, University of Science and
Technology of China, Hefei, Anhui 230026, China }

\begin{abstract}
We study how the mass and magnetic moment of the quarks are dynamically generated in nonequilibrium quark matter.
We derive the equal-time transport and constraint equations for the quark Wigner function in a magnetized quark model
and solve them in the semi-classical expansion. The quark mass and magnetic moment are self-consistently coupled to
the Wigner function and controlled by the kinetic equations. While the quark mass is dynamically generated at the
classical level, the quark magnetic moment is a pure quantum effect, induced by the quark spin interaction
with the external magnetic field.
\end{abstract}

\date{\today}
\maketitle

The intrinsic magnetic moment of an electron is related to its spin by ${\bf m}_s=g \mu_B {\bf s}$, where
$\mu_B=e/(2m)$ is Bohr's magneton, with $e$ and $m$ being the electron charge and mass, $g$ the Lande factor,
and ${\bf s}$ the electron spin angular momentum, respectively. Dirac theory predicts $g=2$ in the non-relativistic limit,
but this result was later challenged by many refined experimental measurements, showing a larger $g$ factor.
Schwinger calculated the first-order radiative correction to ${\bf m}_s$ from the electron-photon
interaction~\cite{schwinger}.
The one-loop contribution to the fermion self-energy was taken into account in a weak magnetic field, which leads to an
anomalous magnetic moment, reflected in a correction to the $g$ factor of order $\sim (g-2)/2=\alpha/(2\pi)$,
where $\alpha$ is the fine-structure constant. Higher-order radiative corrections to $g$ have subsequently been
considered~\cite{high,high1}, resulting in a series in powers of $\alpha/\pi$. These corrections are in excellent
agreement with experimental data. In an external magnetic field ${\bf B}$, the anomalous magnetic moment affects the
electron energy in the lowest Landau level by turning the mass into $m^2_{eff}\simeq m^2+(g-2) eB/2$. In the case of
massless quantum electrodynamics (QED), the anomalous magnetic moment cannot be described through Schwinger's
perturbative approach, and chiral symmetry breaking will dynamically generate an anomalous magnetic
moment~\cite{chiralmag2}.

The anomalous magnetic moment in QED is a fundamental phenomenon in gauge field theory. It
should happen also for quarks in quantum chromodynamics (QCD)~\cite{chiralmag1,chiralmag3}. Considering its
non-Abelian and non-perturbative properties, it becomes much more difficult to directly investigate the quantum
fluctuations in QCD, and effective models without gauge fields, like chiral perturbation theory~\cite{cpt1,cpt2} at the
hadron level and the Nambu--Jona-Lasinio model~\cite{njl1,njl2,njl3,njl4} at the quark level, are used to calculate the
properties of, and spontaneous symmetry breaking in, strong-interaction systems. For instance,
$U(1)_A$ symmetry breaking and spontaneous chiral symmetry breaking in vacuum and their restoration in medium are
investigated for thermal equilibrium systems in an $SU(3)$ linear sigma model~\cite{jiang} and for
non-equilibrium systems
in an NJL model~\cite{guo}. In the chiral limit, the quark magnetic moment is closely related to the chiral symmetry of
QCD~\cite{chiralmag2,chiralmag1,chiralmag3}, which is spontaneously broken in vacuum through the chiral
condensate $\langle \bar\psi \psi\rangle$ or the dynamical quark mass $m_q$. Recent lattice-QCD
simulations~\cite{lattice1,lattice2,lattice3} show that the breaking is further enhanced in an external magnetic field.
Since the constituent quark and anti-quark of the chiral condensate have opposite spins and opposite charges,
the pair's magnetic moment will align with the magnetic field, leading to a
condensate $\langle \bar \psi \gamma_1 \gamma_2 \tau_3 \psi\rangle$ in the ground state. Therefore, the chiral
condensate
will inexorably provide the quasi-particles with both a dynamical mass and a dynamical magnetic moment. The tensor
condensate is discussed at finite temperature in a one-flavor NJL model in the lowest-Landau-level approximation
in a magnetic field in Ref.~\cite{amm2}, and the discussion is extended to a two-flavor NJL model at finite
density in Ref.~\cite{amm3}.

The only possibility to realize a magnetic field in the laboratory which is comparable in strength with typical QCD energy
scales is via high-energy heavy-ion collisions. For heavy-ion collisions at the Relativistic Heavy-Ion Collider and the
Large Hadron Collider, the magnetic field can reach a magnitude of $eB\sim m_\pi^2$~\cite{b2,b3,b4,b5}, however, only
for a very short time in the early stage of the collision. Considering that the colliding system is initially in
a state far from equilibrium, one should study the magnetic moment induced by chiral symmetry breaking
in the framework of quantum transport theory. One
possible way to formulate this theory is the Wigner-function formalism~\cite{qt1,qt2,qt3,qt4,qt5}.
In this Letter, we study the space-time dependent magnetic moment
dynamically generated in quark matter, by applying equal-time transport theory~\cite{qt4,qt5} to an $SU(2)$ NJL model.
We first calculate the temperature dependence of the dynamical magnetic moment in the equilibrium case, and then
focus
on the classical and quantum kinetic equations for the dynamical quark mass and the dynamical magnetic moment.

The Lagrangian of the magnetized $SU(2)$ NJL model with a tensor interaction reads~\cite{njl1,njl2,njl3,njl4}
\begin{equation}
\label{gs}
{\cal L} = \bar{\psi}\left(i\gamma^\mu{\cal D}_\mu-m_0\right)\psi+G_s\left[\left( \bar{\psi} \psi \right)^2
+ \left( \bar{\psi} i \gamma_5 {\bf \tau} \psi \right)^2  \right]-{G_t\over 4}\left[\left( \bar{\psi} \gamma_{\mu} \gamma_{\nu}
{\bf \tau} \psi \right)^2 + \left( \bar{\psi} i \gamma_5 \gamma_{\mu} \gamma_{\nu} \psi \right)^2 \right]\;,
\end{equation}
where the covariant derivative ${\cal D}_\mu=\partial_\mu+iQ A_\mu$ couples quarks with electric charge $Q=
\mathrm{diag} (Q_u,Q_d)=\mathrm{diag} (2e/3,-e/3)$ to an external magnetic field ${\bf B}=(0, 0, B)$ pointing in the
$x_3$-direction through the potential $A_\mu=(0,0,Bx_1,0)$. The coupling constant $G_s$ in the
scalar/pseudo-scalar channel controls the spontaneous chiral symmetry breaking,
which generates a dynamical quark mass, and the coupling constant $G_t$ in the tensor/pseudo-tensor channels
controls the spin-spin interaction, which leads to a dynamical magnetic moment.
Here, $m_0$ is the current quark mass characterizing the explicit chiral symmetry breaking. In the following, we
focus on the chiral limit with $m_0=0$. When the magnetic field is turned on, the chiral symmetry
$SU(2)_L \otimes SU(2)_R$ is reduced to $U(1)_L \otimes U(1)_R$. Throughout the paper we use the notation
$\mathbf{a}=(a_1,a_2,a_3)$ for 3-vectors and $a^\mu = (a_0, \mathbf{a})$ for 4-vectors.

The order parameter for the chiral phase transition is the chiral condensate
$\langle \bar \psi \psi\rangle$ or the dynamical quark mass $m_q=-2G_s \langle \bar \psi \psi\rangle$.
In a magnetic field, we also  introduce a tensor condensate $F_3=-i G_t\langle \bar \psi \gamma_1 \gamma_2
\tau_3 \psi\rangle$, which plays the role of the dynamical magnetic moment of the quarks. Here we consider the
dynamical magnetic moment along the direction of the magnetic field. In mean-field approximation, the Lagrangian
of the model becomes
\begin{eqnarray}
\label{lmf}
{\cal L} = \bar\psi\left(i\gamma^\mu{\cal D}_\mu-m_q-i F_3 \gamma_1\gamma_2 \tau_3\right)\psi-\frac{m_q^2}{4G_s}
-\frac{F_3^2}{2G_t}\;.
\end{eqnarray}
By taking the quark propagator in a magnetic field in the Ritus scheme~\cite{ritus1,ritus2,ritus3}, the thermodynamical
potential of the quark system contains a mean-field part and a quasi-quark part,
\begin{eqnarray}
\label{omega}
\Omega &=&\frac{m_q^2}{4G_s}+\frac{F_3^2}{2G_t}+\Omega_q\;,\\
\Omega_q &=&-N_c\sum_{f,\eta,n}\int \frac{dp_3}{2\pi} \frac{|Q_f B|}{2\pi} \left[\epsilon_{f\eta n}
- 2T\ln g(-\epsilon_{f\eta n})\right]\;,\nonumber
\end{eqnarray}
where $g(x)=\left(1+e^{x/T}\right)^{-1}$ is the Fermi-Dirac distribution, $\epsilon_{f\eta n}=\sqrt{p^2_3
+\left( \sqrt{m_q^2+2n|Q_f B|}+\eta F_3\right)^2}$ is the quark energy of flavor $f=u, d$, and the summation over
the discrete Landau energy levels runs over $n=0,1,2, \ldots$ for $\eta=+$ and over $n=1,2,3, \ldots$ for $\eta=-$.
The spectrum of the quasi-quarks in Landau levels $n>0$ exhibits a Zeeman splitting ($\eta=\pm$) due to the tensor
condensate $F_3$. Therefore, we always use the term ``dynamical magnetic moment'' for the tensor
condensate $F_3$. No splitting is present in the $n=0$ mode, since the fermion in the lowest Landau level has only
one spin projection. The dynamical quark mass and magnetic moment are self-consistently determined by the
minimum of the thermodynamic potential,
\begin{equation}
\label{gap1}
{\partial\Omega\over \partial m_q}=0\;,\ \ \ \ {\partial\Omega\over \partial F_3}=0\;.
\end{equation}

Because of the contact interaction among quarks, the NJL model is non-renormalizable, and it is necessary to introduce a
regularization scheme to remove the ultraviolet divergences of the momentum integrals.
To guarantee the law of causality in
a magnetic field, we take a covariant Pauli-Villars regularization as explained in detail in Ref.~\cite{mao}. The
two parameters of the model in the
chiral limit, namely the quark coupling constant $G_s=3.52$ GeV$^{-2}$ and the Pauli-Villars mass
parameter $\Lambda=1127$ MeV
are fixed by fitting the pion decay constant $f_\pi=93$ MeV and the chiral condensate
$\langle \bar \psi \psi \rangle =(-250\ \text{MeV})^3$ in vacuum at $T=B=0$. The coupling constant $G_t$ in the
tensor channel is treated as a free parameter.

Let us first consider the lowest-Landau-level approximation. In this case, the two gap equations simplify considerably and
become
\begin{eqnarray}
\label{gap2}
&& \frac{m_q}{2G_s}+(m_q+F_3)I_0=0\;,\nonumber\\
&& \frac{F_3}{G_t}+(m_q+F_3)I_0=0\;,
\end{eqnarray}
with
\begin{equation}
I_0=-N_c \frac{|eB|}{\left(2\pi \right)^2} \int \frac{d{p_3}}{\epsilon_3} \left[1-2g\left(\epsilon_3\right)\right]\;.
\end{equation}
The quark energy $\epsilon_{f\eta n}$ becomes flavor-independent in the lowest Landau level with $n=0$ and
$\eta=+$, $\epsilon_3=\sqrt{p^2_3+(m_q+F_3)^2}$.

From the two gap equations (\ref{gap2}), we readily observe that the dynamical
magnetic moment $F_3$ and the dynamical quark mass $m_q$ are proportional to each other,
\begin{equation}
\label{solvemf}
\frac{F_3}{m_q}=\frac{G_t}{2G_s},
\end{equation}
independent of temperature, magnetic field, and the regularization scheme used. Once quarks acquire a
dynamical mass, they should also acquire a dynamical magnetic moment. This effect has also been reported in
massless QED and in a one-flavor NJL model~\cite{chiralmag2,chiralmag1,amm2}. The constituent quark and anti-quark
forming the chiral condensate have opposite spins and opposite charges, the magnetic moment of the pair is then aligned
with the external magnetic field. This leads to a dynamical magnetic moment $F_3$ in the ground state. From the view of
symmetry, once the chiral symmetry is dynamically broken, there is no symmetry protecting the dynamical magnetic
moment, because a nonvanishing value of the latter breaks exactly the same symmetry.

\begin{figure}[hbt]
\centering
\includegraphics[width=8cm]{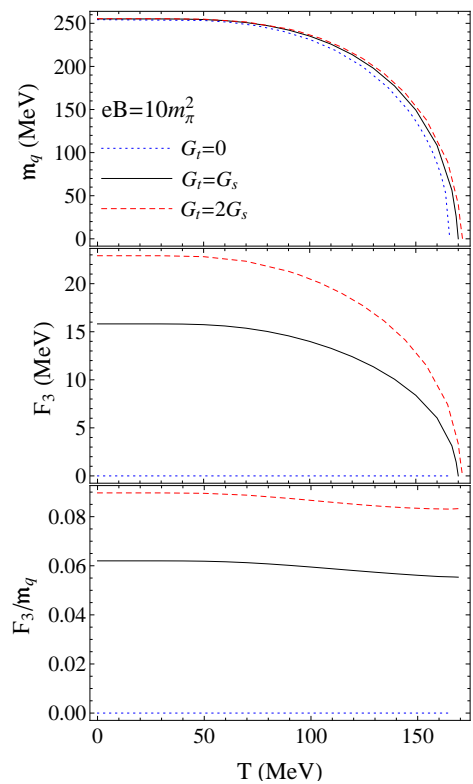}
\caption{ The dynamical quark mass, dynamical magnetic moment, and their ratio as functions of temperature in a
constant magnetic field $eB=10\ m^2_\pi$ for different values of the coupling strength $G_t$ in the tensor channel. }
\label{fig1}
\end{figure}

Including all Landau levels, the proportionality (\ref{solvemf}) between $F_3$ and $m_q$ no longer holds
exactly, but is still
approximately satisfied, see the numerical calculations of the original gap equations (\ref{gap1}) shown in Fig.~\ref{fig1}.
With increasing temperature, the scalar and tensor condensates continuously melt and approach zero at the same critical
temperature, and $F_3$ remains zero in the chirally restored phase, characterized by $m_q=0$.
This proves the original idea that the dynamical magnetic moment is induced by chiral symmetry breaking.
With increasing
coupling strength $G_t$ in the tensor channel, $F_3$ is significantly enhanced but $m_q$ changes only slightly.
While there is still an approximate proportionality between $F_3$ and $m_q$, the proportionality constant in the
full calculation is much smaller than $G_t/(2G_s)$ in the lowest-Landau-level approximation, see the lower panel of
Fig.~\ref{fig1}. This is due to the different contributions from the higher Landau levels to $m_q$ and $F_3$. The quarks in
higher Landau levels participate in the scalar condensate in the same way as the quark in the lowest Landau level and
therefore enhance the dynamical quark mass considerably. However, the quark in the lowest Landau level constitutes
the major contribution to the dynamical magnetic moment due to its single spin projection. Including higher Landau levels,
the dynamical magnetic moment is only slightly changed because of the cancellation between the two spin projections of
the quarks in higher Landau levels.

Apart from the nonzero coupling $G_t$, the other necessary condition for a nonvanishing dynamical magnetic moment is
a nonzero external magnetic field. When the magnetic field is turned off, the gap equations (\ref{gap1}) become
\begin{eqnarray}
\label{gape}
m_q\left\{1+2 G_s N_c N_f \sum_\eta \int \frac{d^3 {\bf p}}{(2 \pi)^3}\frac{1+\eta F_3/\epsilon_\perp}{\epsilon_\eta}
\left[g(\epsilon_\eta)-g(-\epsilon_\eta)\right]\right\}& =& 0\;,\nonumber\\
F_3 + G_t N_c N_f \sum_\eta \int \frac{d^3 {\bf p}}{(2 \pi)^3}\frac{F_3+\eta \epsilon_\perp}{\epsilon_\eta}
\left[g(\epsilon_\eta) -g(-\epsilon_\eta)\right]&=&0\;,
\end{eqnarray}
with quark energy $\epsilon_\eta=\sqrt{p^2_3+(\epsilon_\perp+\eta F_3)^2}$ and transverse energy
$\epsilon_\perp=\sqrt{p^2_1+p^2_2+m_q^2}$. The solution of the gap equations is $F_3=0$ in both the
chiral symmetry broken and restored phases. Physically, without magnetic field, the randomly oriented quark spins
lead to a vanishing dynamical magnetic moment in the ground state.

\begin{figure}[hbt]
\centering
\includegraphics[width=8cm]{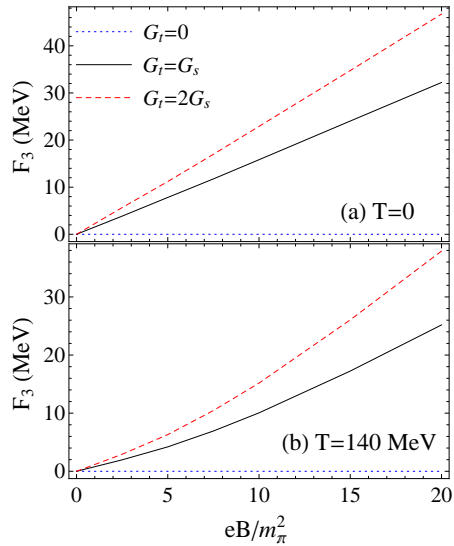}
\caption{ The dynamical magnetic moment as a function of magnetic field at different temperature and different values
for the coupling constant $G_t$ in the tensor channel. }
\label{fig2}
\end{figure}

As the magnetic field is turned on, a nonzero dynamical magnetic moment is induced and increases with magnetic field.
Figure~\ref{fig2} shows the dynamical magnetic moment as a function of magnetic field at zero and finite temperature for
different values of the tensor coupling $G_t$. The dynamical magnetic moment is linearly proportional to the external
magnetic field at zero temperature, analogously to the anomalous magnetic moment in Schwinger's calculation in
QED~\cite{schwinger}. The linear relation is broken by the thermal motion of quarks, see the
lower panel of Fig.~\ref{fig2}.

We now turn to non-equilibrium systems. For systems in a sufficiently strong magnetic field, like matter created
in the early stages of relativistic heavy-ion collisions, the calculation in the framework of finite-temperature field theory
fails, and we need to
treat the dynamical evolution of the system in the framework of transport theory.
In the following, we consider the dynamical
evolution of the quark mass and magnetic moment in an external electromagnetic field by using the Wigner-function
formalism applied to the NJL model with a tensor interaction. To appropriately treat the quantum fluctuations,
especially the
off-shell effect, order by order, we apply equal-time quantum transport theory, which has been successfully developed in
QED~\cite{qt3,qt4,qt5}. We will see clearly that the dynamical quark mass is generated at the
classical level, but the magnetic moment arises from quantum fluctuations.

The covariant quark Wigner function in a gauge field theory is defined as
\begin{equation}
\label{wigner}
W(x,p) = \int d^4 y  e^{ipy}\left\langle \psi(x_+) e^{iQ \int^{1/2}_{-{1/2}} ds A(x+sy)y } \bar{\psi}(x_-) \right\rangle\;,
\end{equation}
where the exponential function is the gauge link between the two points $x_-=x-y/2$ and $x_+=x+y/2$, which guarantees
gauge invariance~\cite{qt3}, and the symbol $\langle ... \rangle$ means ensemble average of the Wigner operator.
For external (classical) gauge fields, the link factor can be moved out of the ensemble average.

From the mean-field Lagrangian (\ref{lmf}) in the chiral limit, we obtain the Dirac equation for the quark field,
\begin{equation}
\left(i\gamma^\mu {\cal D}_\mu -m_q-i F_3 \gamma_1 \gamma_2 \tau_3 \right) \psi=0\;.
\end{equation}
Again, we consider here the dynamical magnetic moment $F_3$ along the direction of the magnetic field.

Using the Dirac equation, we derive the generalized Vasak-Gyulassy-Elze equation~\cite{qt3} for the quark
Wigner function for flavor $f$,
\begin{equation}
\label{cw}
\left(\gamma^\mu K_\mu-M+K_3 \gamma_1\gamma_2 \right) W=0\;,
\end{equation}
with the operators
\begin{eqnarray}
K_\mu &=& \Pi_\mu+\frac{i}{2} \hbar D_{\mu}\;,\nonumber\\
\Pi_\mu &=&p_\mu-i\hbar Q_f  \int^{1/2}_{-1/2} ds s F_{\mu \nu}(x-i\hbar s \partial_p) \partial ^\nu_p\;, \nonumber\\
D_\mu &=& \partial_\mu -Q_f \int^{1/2}_{-1/2} ds   F_{\mu \nu}(x-i\hbar s \partial_p) \partial ^\nu_p\;,
\end{eqnarray}
related to the electromagnetic interaction,
\begin{eqnarray}
M &=& M_1+iM_2\;, \nonumber\\
M_1 &=& \cos\left(\frac{\hbar}{2}\partial_x \cdot \partial_p \right) m_q(x)\;,\nonumber\\
M_2 &=& -\sin\left(\frac{\hbar}{2}\partial_x \cdot \partial_p \right) m_q(x)\;,
\end{eqnarray}
related to the dynamical quark mass controlled by the scalar interaction, and
\begin{eqnarray}
K_3 &=& F_o+iF_e\;, \nonumber\\
F_e &=& -\text{sgn}(Q_f)\cos\left(\frac{\hbar}{2}\partial_x \cdot \partial_p \right) F_3(x)\;,\nonumber\\
F_o &=&- \text{sgn}(Q_f)\sin\left(\frac{\hbar}{2}\partial_x \cdot \partial_p \right) F_3(x)\;,
\end{eqnarray}
related to the dynamical magnetic moment controlled by the tensor interaction. We have explicitly exhibited the
$\hbar$-dependence in order to be able to discuss the semi-classical expansion of the kinetic equation in the following.
Considering that the Wigner function defined through Eq.~(\ref{wigner}) is a $4\times 4$ matrix in Dirac space and
in general not a real-valued function, its physical meaning becomes clear only after the spinor decomposition~\cite{qt3}
\begin{equation}
\label{spin4}
W = \frac{1}{4}\left(F+i \gamma_5 P+\gamma_\mu V^\mu+\gamma_\mu \gamma_5 A^\mu
+\frac{1}{2}\sigma_{\mu \nu} S^{\mu \nu} \right)\;.
\end{equation}

To compare the covariant Wigner function $W(x,p)$ defined in $4-$dimensional momentum space with the observable
physics densities such as the number density defined in $3-$dimensional momentum space, we introduce the equal-time
Wigner function ${\cal W}(x,{\bf p})$ by integrating the covariant Wigner function $W(x,p)$ over the energy $p_0$ and
furthermore apply the corresponding spinor decomposition,
\begin{eqnarray}
\label{spin3}
{\cal W} &=& \int dp_0 W \gamma_0\nonumber\\
&=&\frac{1}{4}\left(f_0+\gamma_5 f_1-i\gamma_0 \gamma_5 f_2 +\gamma_0 f_3+\gamma_5 \gamma_0 {\bf \gamma}
\cdot {\bf g}_0+\gamma_0 {\bf \gamma} \cdot {\bf g}_1-i{\bf \gamma} \cdot {\bf g}_2-\gamma_5 {\bf \gamma}
\cdot {\bf g}_3\right)\;.
\end{eqnarray}
The physical meaning of the spinor components of the equal-time Wigner function $f_i(x,{\bf p})$ and ${\bf g}_i(x,{\bf p})$,
$i=0,1,2,3$, is discussed in detail in Ref.~\cite{qt4} in QED. For instance, $f_0$ is the number
density, ${\bf g}_0$ the spin density, and ${\bf g}_1$ the number current.

Since the kinetic equation (\ref{cw}) is a complete equation, when taking the spinor decomposition (\ref{spin4}) it
becomes $16$ transport equations with derivative $D_\mu$ plus $16$ constraint equations with operator $\Pi_0$ for
the spinor components $F, P, V^\mu, A^\mu$, and $S^{\mu\nu}$. The former controls the dynamical evolution of the
$16$ components in phase space, and the latter is the quantum extension of the classical on-shell condition~\cite{qt5}.
By taking the energy integration of these kinetic equations, we obtain a set of transport equations for the spinor
components of the equal-time Wigner function,
\begin{eqnarray}
\label{transport3}
&& {\hbar\over 2}\left(d_0 f_0+ {\bf d} \cdot {\bf g}_1 \right)-m_2f_3-f_o {\bf g}_3\cdot{\bf e}_3 =0\;,\nonumber\\
&& {\hbar\over 2}\left(d_0 f_1+ {\bf d} \cdot {\bf g}_0 \right)+m_1f_2-f_e {\bf g}_2\cdot{\bf e}_3 =0\;, \nonumber\\
&& {\hbar\over 2} d_0 f_2 +{\bf \pi} \cdot {\bf g}_3 -m_1 f_1 +f_e {\bf g}_1\cdot{\bf e}_3 =0\;,\nonumber\\
&& {\hbar\over 2} d_0 f_3 -{\bf \pi} \cdot {\bf g}_2 -m_2 f_0 -f_o {\bf g}_0\cdot{\bf e}_3 =0\;,\nonumber\\
&& {\hbar\over 2}\left(d_0 {\bf g}_0 +{\bf d}f_1 \right)-{\bf \pi}\times {\bf g}_1-m_2{\bf g}_3 -f_e{\bf g}_3\times{\bf e}_3
-f_of_3\mathbf{e}_3=0\;,\nonumber\\
&& {\hbar\over 2}\left(d_0 {\bf g}_1 +{\bf d}f_0 \right)-{\bf \pi}\times {\bf g}_0+m_1{\bf g}_2 +f_o{\bf g}_2\times{\bf e}_3
-f_ef_2\mathbf{e}_3=0\;,\nonumber\\
&& {\hbar\over 2}\left(d_0 {\bf g}_2 +{\bf d}\times{\bf g}_3\right)+{\bf \pi}f_3-m_1{\bf g}_1 -f_o{\bf g}_1\times{\bf e}_3
+f_ef_1\mathbf{e}_3=0\;,\nonumber\\
&& {\hbar\over 2}\left(d_0 {\bf g}_3 -{\bf d}\times{\bf g}_2\right)-{\bf \pi}f_2-m_2{\bf g}_0 -f_e{\bf g}_0\times{\bf e}_3
-f_of_0\mathbf{e}_3=0\;,
\end{eqnarray}
and a set of constraint equations,
\begin{eqnarray}
\label{constraint3}
&& V_0' + \pi_0 f_0- {\bf \pi} \cdot {\bf g}_1-m_1 f_3+f_e {\bf g}_3\cdot{\bf e}_3=0\;,\nonumber\\
&& A_0' - \pi_0 f_1+ {\bf \pi} \cdot {\bf g}_0+m_2 f_2+f_o {\bf g}_2\cdot{\bf e}_3=0\;,\nonumber\\
&& P' + \pi_0 f_2+ {\hbar\over 2}{\bf d}\cdot{\bf g}_3+m_2 f_1+f_o {\bf g}_1\cdot{\bf e}_3=0\;,\nonumber\\
&& F' + \pi_0 f_3- {\hbar\over 2}{\bf d}\cdot{\bf g}_2-m_1 f_0+f_e {\bf g}_0\cdot{\bf e}_3=0\;,\nonumber\\
&& {\bf A}' - \pi_0 {\bf g}_0+ {\hbar\over 2} {\bf d} \times {\bf g}_1+{\bf \pi} f_1+m_1 {\bf g}_3+f_o{\bf g}_3\times{\bf e}_3
-f_ef_3\mathbf{e}_3=0\;,\nonumber\\
&& {\bf V}' + \pi_0 {\bf g}_1- {\hbar\over 2} {\bf d} \times {\bf g}_0-{\bf \pi} f_0-m_2 {\bf g}_2-f_e{\bf g}_2\times{\bf e}_3
-f_of_2\mathbf{e}_3=0\;,\nonumber\\
&& S_{0i}'{\bf e}_i - \pi_0 {\bf g}_2+{\bf \pi}\times {\bf g}_3- {\hbar\over 2} {\bf d} f_3-m_2 {\bf g}_1
-f_e{\bf g}_1\times{\bf e}_3
-f_of_1\mathbf{e}_3=0\;,\nonumber\\
&& S_{jk}'\epsilon^{ijk}{\bf e}_i +2 \pi_0 {\bf g}_3+2{\bf \pi}\times {\bf g}_2- \hbar {\bf d} f_2-2m_1 {\bf g}_0-f_o{\bf g}_0
\times{\bf e}_3+f_ef_0\mathbf{e}_3=0\;,
\end{eqnarray}
where $\Gamma'(x,{\bf p})=\int dp_0 p_0\Gamma(x,p)\ (\Gamma=F,P,V^\mu,A^\mu,S^{\mu\nu})$
are the first-order energy
moments of the covariant Wigner function, ${\bf e}_1, {\bf e}_2$, and ${\bf e}_3$ are the unit vectors along the
Cartesian coordinates $x_1, x_2$, and $x_3$ in coordinate space, and the equal-time operators related to the quark
electromagnetic, scalar, and tensor interactions are the energy integrals of the corresponding covariant operators,
\begin{eqnarray}
\label{operator3}
d_0 &=& \partial_t+Q_f \int^{1/2}_{-1/2} ds \ {\bf E}({\bf x}+i\hbar s{\bf \nabla}_p,t) \cdot {\bf \nabla}_p\;,\nonumber\\
{\bf d} &=& {\bf \nabla}+Q_f \int^{1/2}_{-1/2} ds \ {\bf B}({\bf x}+i\hbar s{\bf \nabla}_p,t) \times {\bf \nabla}_p\;,\nonumber\\
\pi_0 &=& i \hbar Q_f  \int^{1/2}_{-1/2} ds \ s {\bf E}({\bf x}+i\hbar s{\bf \nabla}_p,t) \cdot {\bf \nabla}_p\;,\nonumber\\
{\bf \pi} &=& {\bf p} - i \hbar Q_f  \int^{1/2}_{-1/2} ds \ s{\bf B}({\bf x}+i\hbar s{\bf \nabla}_p,t) \times {\bf \nabla}_p\;,
\nonumber\\
m_1 &=& \cos\left({\hbar\over 2} {\bf \nabla} \cdot {\bf \nabla}_p\right) m_q(x)\;,\nonumber\\
m_2 &=& \sin\left({\hbar\over 2} {\bf \nabla} \cdot {\bf \nabla}_p\right) m_q(x)\;,\nonumber\\
f_e &=& -\text{sgn}(Q_f)\cos\left({\hbar\over 2} {\bf \nabla} \cdot {\bf \nabla}_p\right) F_3(x)\;,\nonumber\\
f_o &=& \text{sgn}(Q_f)\sin\left({\hbar\over 2} {\bf \nabla} \cdot {\bf \nabla}_p\right) F_3(x)\;.
\end{eqnarray}
Here we have replaced the field strength tensor $F_{\mu\nu}(x)$ by the electric and magnetic fields ${\bf E}(x)$
and ${\bf B}(x)$.
Note that the energy moment $\int dp_0 p_0 W(x,p)\gamma_0$ in the constraint equations is in general
independent of the equal-time Wigner function ${\cal W}(x,{\bf p})$ due to the quantum off-shell effect of particle transport
in the medium~\cite{qt5}. Only in the classical case, any order energy moment can be expressed as $\int dp_0 p_0^n
W(x,p)\gamma_0=\omega_{\bf p}^n {\cal W}(x,{\bf p}),\, n=0,1,2,\ldots$, in terms of the quasi-particle
energy $\omega_{\bf p}$ and the equal-time Wigner function due to the classical on-shell condition
$\delta(p_0-\omega_{\bf p})$.

Using the definitions of the scalar and tensor condensates $m_q=-2G_s \langle \bar \psi \psi \rangle
=-2G_s\langle\bar\psi_u\psi_u+\bar\psi_d\psi_d\rangle$ and $F_3=-i G_t \langle \bar \psi \gamma_1 \gamma_2 \tau_3
\psi \rangle=-i G_t \langle \bar\psi_u\gamma_1\gamma_2\psi_u-\bar\psi_d\gamma_1\gamma_2\psi_d \rangle$, these
quantities can be expressed in terms of the Wigner function,
\begin{eqnarray}
\label{mf}
&&m_q(x)=-2G_s \int \frac{d^3{\bf p}}{(2\pi)^3} \left[f_{3u}(x,{\bf p}) +f_{3d}(x,{\bf p}) \right]\;, \nonumber\\
&&F_3(x)=-G_t \int \frac{d^3{\bf p}}{(2\pi)^3} \left[{\bf g}_{3u}(x,{\bf p}) -{\bf g}_{3d}(x,{\bf p}) \right]\cdot{\bf e}_3\;.
\end{eqnarray}
This shows clearly the physics of the spinor components $f_3$ and ${\bf g}_3$: they are the source of the quark mass
and the quark magnetic moment, respectively, and are then called mass density and magnetic-moment density.
By solving the kinetic equations (\ref{transport3}) and (\ref{constraint3}), the quark mass and magnetic moment
are self-consistently generated through the dynamical evolution of the quark Wigner function.

To see clearly the quantum effect on the equal-time kinetic theory, we apply the semi-classical ($\hbar$) expansion
 for the Wigner functions and the equal-time operators,
\begin{eqnarray}
\label{hbar}
W &=& W^{(0)}+\hbar W^{(1)}+{\cal O}(\hbar^2)\;,\nonumber\\
{\cal W} &=& {\cal W}^{(0)}+\hbar{\cal W}^{(1)}+{\cal O}(\hbar^2)\;,\nonumber\\
d_0 &=& \partial_t+Q_f{\bf E}\cdot{\bf \nabla}_p+{\cal O}(\hbar^2)\;,\nonumber\\
{\bf d} &=& {\bf \nabla}+Q_f{\bf B}\times {\bf \nabla}_p+{\cal O}(\hbar^2)\;,\nonumber\\
\pi_0 &=& {\cal O}(\hbar^2)\;,\nonumber\\
{\bf \pi} &=& {\bf p}+{\cal O}(\hbar^2)\;,\nonumber\\
m_1 &=& m_q+{\cal O}(\hbar)\;,\nonumber\\
m_2 &=& -{\hbar\over 2}{\bf \nabla}m_q\cdot{\bf \nabla}_p+{\cal O}(\hbar^2)\;,\nonumber\\
f_e &=& -\text{sgn}(Q_f)F_3+{\cal O}(\hbar)\;,\nonumber\\
f_o &=& -{\hbar\over 2}\text{sgn}(Q_f){\bf \nabla}F_3\cdot{\bf \nabla}_p+{\cal O}(\hbar^2)\;.
\end{eqnarray}
By substituting them into the kinetic equations and comparing orders of $\hbar$ on both sides, we obtain the transport
and constraint equations order by order in $\hbar$. In the classical limit, i.e., $\hbar=0$, the constraint equations
(\ref{constraint3}) determine automatically the on-shell energy
\begin{equation}
p_0=\chi \epsilon_\eta\;,\ \ \ \ \ \chi,\;\eta=\pm\;,
\end{equation}
corresponding to the four independent quasi-particle solutions with positive and negative energies ($\chi=\pm$) and up
and down spin projections ($\eta=\pm$). In this case we can express the distribution functions as the sum of
the distributions for the four quasi-particle modes,
$f_i=\sum_{\chi,\eta}f_i^{\chi\eta}$ and ${\bf g}_i=\sum_{\chi,\eta}{\bf g}_i^{\chi\eta}$.
To simplify the notation, we have here and in the following neglected the subscript $(0)$ of the classical components
$f_i^{(0)}$ and ${\bf g}_i^{(0)}$. The constraint equations determine not only the on-shell condition but also give rise
to relations among the classical components,
\begin{eqnarray}
\label{constraint30}
f_1^{\chi\eta} &=& \text{sgn}(Q_f)\ \chi\eta{m_q\over \epsilon_\eta}{ p_3\over \epsilon_\perp}f_0^{\chi\eta}\;,
\nonumber\\
f_2^{\chi\eta} &=& 0\;, \nonumber\\
f_3^{\chi\eta} &=& \chi{m_q\over \epsilon_\eta}\left(1+\eta{F_3\over \epsilon_\perp}\right)f_0^{\chi\eta}\;,\nonumber\\
{\bf g}_0^{\chi\eta} &=& \text{sgn}(Q_f)\ \eta {m_q\over \epsilon_\perp}\mathbf{e}_3 f_0^{\chi\eta}\;,
\nonumber\\
{\bf g}_1^{\chi\eta} &=& \chi {1\over \epsilon_\eta}\left[{\bf p}-\eta{F_3\over
\epsilon_\perp}({\bf p}\times{\bf e}_3)\times{\bf e}_3\right]f_0^{\chi\eta}\;, \nonumber\\
{\bf g}_2^{\chi\eta} &=& \text{sgn}(Q_f )\ \eta {{\bf p}\times {\bf e}_3\over \epsilon_\perp}f_0^{\chi\eta}\;, \nonumber\\
{\bf g}_3^{\chi\eta} &=& -\text{sgn}(Q_f)\ \chi{\eta\over \epsilon_\eta\epsilon_\perp}\left[ p_3 {\bf p}
-\left(\epsilon_\perp^2+\eta F_3\epsilon_\perp\right)\mathbf{e}_3\right]f_0^{\chi\eta}\;.
\end{eqnarray}
These relations greatly simplify the calculation of the classical Wigner function. The nonzero tensor condensate couples
the spin-related distributions to the number density-related distributions. Therefore, there is only one independent
distribution function, the number density $f_0$, and all others can be expressed in terms of $f_0$. Note that the
classical limit of the transport equations (\ref{transport3}) can reproduce a part of the classical relations shown in
Eq.~(\ref{constraint30}) but does not give any new relations.

Substituting the classical relations between $f_3, {\bf g}_3$, and $f_0$ into the expressions (\ref{mf}) for $m_q$
and $F_3$,
and considering the trivial color degrees of freedom in the NJL model, the non-trivial quark mass $m_q(x)$ and magnetic
moment $F_3(x)$ at the classical level are controlled by the gap equations
\begin{eqnarray}
\label{gapc}
&& 1+2 G_s N_c \sum_{\chi,\eta} \int {d^3 {\bf p}\over (2 \pi)^3}{1+\eta F_3/\epsilon_\perp\over \chi\epsilon_\eta}
\left(f_{0u}^{\chi\eta}+f_{0d}^{\chi\eta}\right)=0\;,\nonumber\\
&& F_3+G_t N_c \sum_{\chi,\eta} \int {d^3 {\bf p}\over (2 \pi)^3}{F_3+\eta \epsilon_\perp\over \chi\epsilon_\eta}
\left(f_{0u}^{\chi\eta}+f_{0d}^{\chi\eta}\right)=0\;.
\end{eqnarray}
These two classical gap equations have the same structure as Eq.~(\ref{gape}) for systems in thermal equilibrium,
the only difference being the non-equilibrium distribution $f_0(x,{\bf p})$, which is controlled by a classical transport
equation and will be discussed below. When replacing $f_0$ by the Fermi-Dirac distribution, the gap equations
(\ref{gapc}) and (\ref{gape}) become exactly the same. Remember that $F_3=0$ is the only solution of the gap
equations (\ref{gape}), the same structure, namely the same dynamics of Eqs.~(\ref{gapc}) and (\ref{gape}) leads to the
conclusion that $F_3=0$ in the classical limit. Physically, the dynamical magnetic moment $F_3$ is generated by the
quark
spin, which is a quantum phenomenon and will not appear at the classical level. $F_3$ has a nonzero value only when
quantum fluctuations are included. On the other hand, the chiral symmetry restoration in medium is a classical phase
transition, the space-time dependence of the order parameter $m_q(x)$ is solely controlled by
\begin{equation}
1+2 G_s N_c \sum_{\chi,\eta} \int {d^3 {\bf p}\over (2 \pi)^3}{1\over \chi\epsilon_\eta}
\left(f_{0u}^{\chi\eta}+f_{0d}^{\chi\eta}\right)=0\;.
\end{equation}

When the tensor condensate vanishes in the classical limit, the spin density ${\bf g}_0$ becomes an independent
Wigner component, and the constraint equations (\ref{constraint3}) result in the classical relations
\begin{eqnarray}
\label{class9}
f_1^\chi &=& \chi{{\bf p} \cdot {\bf g}^\chi_0\over \epsilon}\;,\nonumber \\
f_2^\chi &=& 0\;,\nonumber\\
f_3^\chi &=& \chi{m_q\over\epsilon} f_0^\chi\;,\nonumber \\
{\bf g}^\chi_1 &=& \chi {{\bf p}\over \epsilon} f_0^\chi\;,\nonumber\\
{\bf g}^\chi_2 &=& {{\bf p} \times {\bf g}^\chi_0\over m_q}\;,\nonumber \\
{\bf g}^\chi_3 &=& \chi{\epsilon^2 {\bf g}^\chi_0 -\left({\bf p} \cdot {\bf g}^\chi_0 \right){\bf p}\over m_q\epsilon}\;,
\end{eqnarray}
with the quark energy $\epsilon=\sqrt{m_q^2+{\bf p}^2}$.

We now consider the transport equations to linear order in $\hbar$. Taking into account the classical solution
$F_3=0$ and $f_2=0$, we have
\begin{eqnarray}
\label{transporthbar}
&& d_0 f_0 + {\bf d} \cdot {\bf g}_1 -{\bf \nabla}m_q\cdot{\bf \nabla}_p f_3=0\;,\nonumber\\
&& d_0 f_1 + {\bf d} \cdot {\bf g}_0+2m_q f_2^{(1)}+2\text{sgn}(Q_f)F_3^{(1)}{\bf g}_2\cdot{\bf e}_3=0\;, \nonumber\\
&& {\bf p}\cdot{\bf g}_3^{(1)}-m_qf_1^{(1)}-m_q^{(1)}f_1-\text{sgn}(Q_f)F_3^{(1)}{\bf g}_1\cdot{\bf e}_3=0\;,\nonumber\\
&& d_0 f_3-2{\bf p} \cdot {\bf g}_2^{(1)}-{\bf \nabla}m_q\cdot{\bf \nabla}_p f_0=0\;, \nonumber\\
&& d_0 {\bf g}_0 +{\bf d} f_1-2{\bf p} \times {\bf g}_1^{(1)}-{\bf \nabla}m_q\cdot{\bf \nabla}_p{\bf g}_3
+2\text{sgn}(Q_f) F_3^{(1)}{\bf g}_3\times{\bf e}_3=0\;, \nonumber \\
&& d_0 {\bf g}_1 +{\bf d} f_0-2{\bf p} \times {\bf g}_0^{(1)}+2m_q {\bf g}_2^{(1)}+2m_q^{(1)}{\bf g}_2=0\;, \nonumber\\
&& d_0 {\bf g}_2+ {\bf d} \times {\bf g}_3+2{\bf p} f_3^{(1)}-2m_q {\bf g}_1^{(1)}-2m_q^{(1)}{\bf g}_1
-2\text{sgn}(Q_f)F_3^{(1)}f_1\mathbf{e}_3=0\;, \nonumber\\
&& d_0 {\bf g}_3 -{\bf d} \times {\bf g}_2-2{\bf p} f_2^{(1)}-{\bf \nabla}m_q\cdot{\bf \nabla}_p{\bf g}_0
+2\text{sgn}(Q_f)F_3^{(1)}{\bf g}_0\times{\bf e}_3=0\;,
\end{eqnarray}
where $f_i^{(1)}, {\bf g}_i^{(1)}, m_q^{(1)}$, and $F_3^{(1)}$ are the first-order quantum corrections, and $d_0$ and
${\bf d}$ are at the classical level, $d_0 =\partial_t+ Q_f {\bf E} \cdot {\bf \nabla}_p$ and
${\bf d} = {\bf \nabla}+Q_f {\bf B} \times {\bf \nabla}_p$.

With the help of the classical relations (\ref{class9}), a careful but straightforward treatment of Eqs.\
(\ref{transporthbar}) determines the quantum correction from the quark spin density ${\bf g}_0$ to the magnetic
moment $F_3^{(1)}$,
{\begin{equation}
\label{quantum1}
2\left(m_q{\bf g}_3+f_1 {\bf p} \right)\cdot{\bf e}_2  \; \text{sgn}(Q_f) F_3^{(1)}
=\left[\frac{{\bf \nabla}m_q^2}{2}\cdot{\bf \nabla}_p{\bf g}_3-m_q\left(d_0{\bf g}_0+{\bf d}f_1\right)
+{\bf p}\times \left(d_0{\bf g}_2+{\bf d}\times{\bf g}_3\right)\right]\cdot{\bf e}_1\;,
\end{equation}
and leads to the transport equations for the two independent classical components, the number density $f_0$ and spin
density ${\bf g}_0$,
\begin{eqnarray}
\label{g0f0}
\left(d_0+\chi{{\bf p}\over {\epsilon}} \cdot {\bf d}-\chi{{\bf \nabla} m_q^2 \cdot {\bf \nabla}_p\over 2\epsilon}\right)f_0^\chi
&=& 0\;,\nonumber\\
\left(d_0+\chi{{\bf p}\over\epsilon} \cdot {\bf d}-\chi{{\bf \nabla} m_q^2 \cdot {\bf \nabla}_p\over 2\epsilon}\right)
{\bf g}_0^\chi
&=& {Q_f\over\epsilon^2}\left[{\bf p} \times \left({\bf E} \times {\bf g}_0^\chi \right) -\chi\epsilon {\bf B} \times
{\bf g}_0^\chi\right]-{1\over 2\epsilon^2 m_q^2}\left(\partial_t m_q^2 {\bf p}+\chi\epsilon{\bf \nabla}m_q^2\right)
\times ({\bf p} \times {\bf g}_0^\chi)\nonumber\\
&&-{{\text{sgn}(Q_f) \chi }\over m_q\epsilon}\left[m_q^2{\bf g}_0^\chi\times{\bf e}_3+\left(({\bf p}\times{\bf g}_0^\chi)
\cdot{\bf e}_3\right){\bf p}\right]F_3^{(1)}\;.
\end{eqnarray}
The quarks obtain a dynamical mass $m_q$ from the interaction with the medium. When the medium is inhomogeneous,
a mean-field force ${\bf F}=-{\bf \nabla}m_q^2/(2\epsilon)$ is exerted on the moving quark, which leads to the third term
on the left-hand side of the two transport equations. While in mean-field approximation there is no collision term on the
right-hand side of the transport equation for the number density $f_0$, the quark spin interactions with the
electromagnetic field, the space-time dependent quark mass, and the magnetic moment lead to the three kinds of
collision terms shown on the right-hand side of the transport equation for the quark spin density ${\bf g}_0$.

Let us now consider the limit of a homogeneous medium and a constant magnetic field. In this limit, the quantum correction
$F_3^{(1)}$ becomes
\begin{equation}
F_3^{(1)} = \frac{|Q_f|\left\{{\bf p}\times\left[({\bf B}\times{\bf \nabla}_p)
\times{\bf g}_3\right]-m_q{\bf B}\times{\bf \nabla}_p f_1\right\}\cdot{\bf e}_1}{2\left(m_q{\bf g}_3
+f_1 {\bf p} \right)\cdot{\bf e}_2}\;.
\end{equation}
It is clear that the quantum correction vanishes when the magnetic field disappears.
Moreover, the quantum correction $F_3^{(1)}$ cannot be generated from the inhomogeneous medium, if the
electromagnetic field is turned off. Substituting the kinetic equation (\ref{g0f0}) for
${\bf g}_0$  into the first-order quantum correction (\ref{quantum1}) to $F_3^{(1)}$, and making use of
the relations in Eq.~(\ref{class9}), we can straightforwardly prove $F_3^{(1)}=0$.

Our strategy to extract quantum corrections from a general kinetic theory is the following. The classical kinetic theory for
quasi-particles arises from the constraint equations at zeroth order in $\hbar$ and the transport equations at first order in
$\hbar$. The quantum correction induced by the spin of the quasi-particles comes also from the transport equations at
first order in $\hbar$. When we go to higher-order quantum corrections, the particles are no longer on the energy shell,
and the quasi-particle treatment fails. In this case, the first-order energy moment $\int dp_0 p_0 W(x,p)\gamma_0$ is
independent of the zeroth-order energy moment $\int dp_0 W(x,p)\gamma_0={\cal W}(x,{\bf p})$. Therefore, all 16 spin
components $f_i^{(j)}$ and ${\bf g}_i^{(j)}$ ($i=0,1,2,3,j=1,2,\ldots$) become independent of each other, and their
behavior is controlled by the full set of transport equations (\ref{transport3}).

In summary, we investigated the dynamically generated quark mass and magnetic moment in the Wigner-function
formalism. We derived the transport and constraint equations for the spinor components of the equal-time Wigner function
in the magnetized NJL model with tensor interaction. We expanded the kinetic equations in the semi-classical expansion
and solved them order by order. The space-time dependent quark mass and magnetic moment are self-consistently coupled
to the Wigner function and determined by the kinetic equations. While the quark mass can be dynamically generated at
the classical level, the quark magnetic moment is induced by quantum fluctuations, namely by the quark spin interaction
with the external magnetic field. \\

\noindent {\bf Acknowledgements} \\
The work of S.J.M.\ is supported by NSFC Grant 11775165. She acknowledges partial support by the
``Extreme Matter Institute'' EMMI funded by the Helmholtz Association. The work of D.H.R.\ is supported by the
Deutsche Forschungsgemeinschaft (DFG, German Research Foundation)
through the Collaborative Research Center CRC-TR 211 ``Strong-interaction matter
under extreme conditions'' -- project number 315477589 - TRR 211. He also acknowledges partial support
by the High-end Foreign Experts project GDW20167100136 of the State
Administration of Foreign Experts Affairs of China.

\end{document}